\documentclass[twoside,a4paper,11pt]{sea22}
\usepackage{graphicx}
\usepackage{hyperref}
\usepackage{media9}
\usepackage{bm}
\begin{document}
\pagenumbering{arabic}
\pagestyle{myheadings}
\thispagestyle{empty}
{\flushleft\includegraphics[width=\textwidth,bb=58 650 590 680]{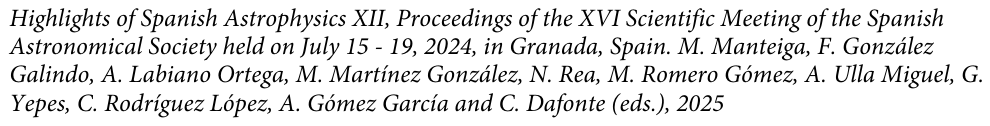}}
\vspace*{0.2cm}
\begin{flushleft}
{\bf {\LARGE
%
Vortex Flows in the Solar Atmosphere: Detection and Heating Mechanisms in 3D MHD Numerical Simulations 
}\\
\vspace*{1cm}
%
Koll Pistarini, M.$^{1,2}$,
Khomenko, E.$^{1,2}$ and 
Felipe, T.$^{1,2}$
%
}\\
\vspace*{0.5cm}
%
$^{1}$
Instituto de Astrofísica de Canarias, 38205 La Laguna, Tenerife, Spain
\\
$^{2}$
Departamento de Astrofísica, Universidad de La Laguna, 38205, La Laguna, Tenerife, Spain\\

%
\end{flushleft}
%
\markboth{
Vortex Flows in the Solar Atmosphere
}{ 
%
Koll Pistarini, M. et al. 
%
}
\thispagestyle{empty}
\vspace*{0.4cm}
\begin{minipage}[l]{0.09\textwidth}
\ 
\end{minipage}
\begin{minipage}[r]{0.9\textwidth}
\vspace{1cm}
\section*{Abstract}{\small
%

Vortex flows are structures associated with the rotation of the plasma and/or the magnetic field that are present throughout the solar atmosphere. In recent years, their study has become increasingly important, as they are present on a wide variety of temporal and spatial scales and can connect several layers of the solar atmosphere. In this work, we focused on the detection and analysis of these structures in an automatic way. We use realistic $3$D MHD numerical simulations obtained with the Mancha3D code at different magnetic field configurations and spatial resolutions. The vortex detection has been performed using the novel SWIRL code. We have been able to determine multiple structures associated with small and large scale vortices that extend in height in our simulations. We performed a statistical analysis of these structures, quantifying their number and typical sizes, as well as their temperature and heating profiles, confirming their importance in the energy transport.

%
\normalsize}
\end{minipage}
%
%
%
\section{Introduction \label{intro}}

In recent years, the study of vortices in the solar atmosphere has become increasingly important \cite{tziotziou23}. It has been shown that these structures can connect and transport energy throughout the different layers of the solar atmosphere \cite{wedemeyer12}. Thus, they have been proposed as an important key to understand the heating of the outer layers of the solar atmosphere.

A vortex can be defined as ``the collective motion associated with the azimuthal component of a vector field (e.g. velocity or magnetic field) about a common centre or axis and that it is persistent in time'' \cite{tziotziou23}. There are evidences of the presence of vortices in both observations \cite{bonet08, wedemeyer09, dakanalis22} and numerical simulations \cite{moll12, silva20, yadav20}. Nowadays, several studies of vortices in numerical simulations have been carried out. Each of them have been done using different numerical codes with a specific magnetic field configuration and spatial resolution. In this work we address the study of energy transport and dissipation mechanisms that take place in vortices under different magnetic field configurations and spatial resolutions using the same numerical code. This will allow us to compare all of our models under the same numerical framework, without the possible influence of using different numerical schemes.

\begin{figure}[!t]
\center
\includegraphics[width=13cm]{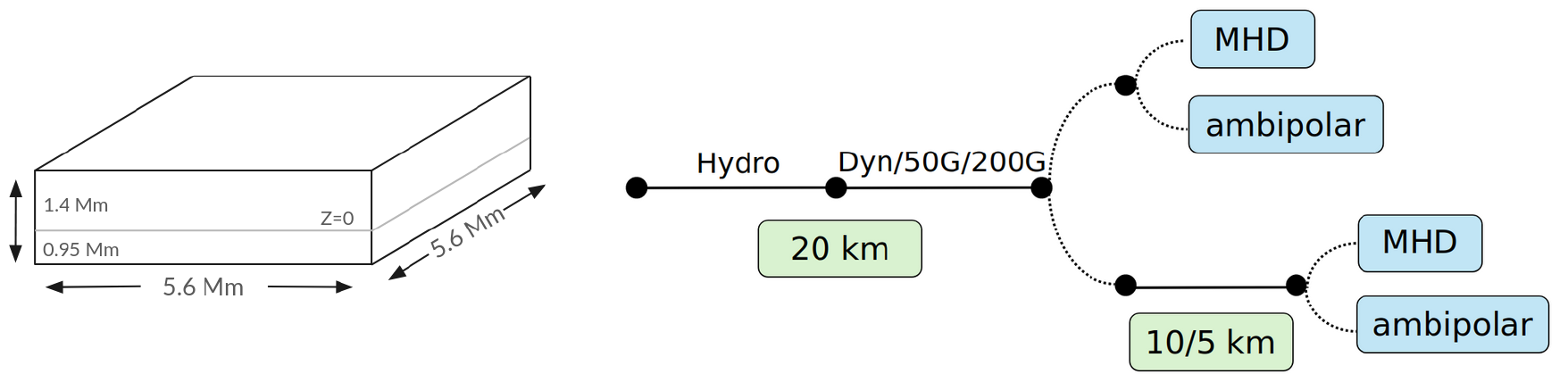}
\caption{\label{fig1} Left side: Illustration of the physical domain used in all the simulations. Right side: schematic representation of the path followed to calculate the different numerical simulations. The blue panels indicate if the dataset includes ambipolar diffusion (ambipolar) or not (MHD). The green panels show the horizontal spatial resolution at each stage.}
\end{figure}

\section{Methodology \label{method}}

\subsection{Numerical simulations \label{sims}}

To perform the analysis of vortices in numerical simulations, we used $3$D magnetoconvection numerical simulations which have been obtained through the Mancha3D code \cite{modestov24}. We have three different magnetic field configurations: small-scale dynamo (SSD) simulation, which are initiated through the Biermann battery term \cite{khomenko17}; and two simulations with an artificially implanted vertical magnetic field of $50$ G and $200$ G. Each magnetic field model is available at different spatial resolutions: $20\times20\times14$ km$^3$ for the three magnetic field configurations, $10\times10\times7$ km$^3$ for the SSD and $50$ G simulations, and $5\times5\times3.5$ km$^3$ just for the SSD simulation. All of the models cover the same physical domain of $5.76\times 5.76\times2.3$ Mm$^3$, expanding approximately $0.95$ Mm below the solar surface and $1.4$ Mm above it. A sketch of the simulation box can be found in the left side of Fig.~\ref{fig1}. Simulations are computed based on the process illustrated on the right side of Fig.~\ref{fig1}. Two similar datasets are available for each model: one that does not include non-ideal MHD effects and one that incorporates ambipolar diffusion. Both datasets evolve from the same initial time instant until about $20$ minutes of solar time, whereafter the two datasets start to be different enough and cannot be compared one to one. The output temporal cadence is $10$ s. 

\subsection{Vortex detection \label{vortex}}

Vortex detection in our simulations has been carried out using the SWIRL code \cite{canivete22}. In particular, we have focused on detecting vortices using the velocity field. The code uses the mathematical quantity Rortex, which is based on the velocity gradient tensor $\mathcal{U}_{ij} := \partial_{j} v_{i}$, as the mathematical background for detecting regions with a rotating velocity field:
\begin{equation}
    R = \bm{\omega} \cdot \bm{u_{r}} - \sqrt{ \left( \bm{\omega} \cdot \bm{u_{r}} \right)^{2} - \lambda^{2} } \; , 
\end{equation}
where $\bm{\omega}=\nabla \times \bm{v} $ is the vorticity, and $\bm{u_{r}}$ and $\lambda$ are the real eigenvector and the eigenvalue associated with the imaginary eigenvector of the velocity gradient tensor $\mathcal{U}$, respectively. Combined with this mathematical criteria, the code includes a clustering algorithm to automatically detect regions associated with vortices. For this purpose, it searches for clusters of the Rortex quantity to determine the presence of vortices. 

Before applying the SWIRL code, we have filtered the simulations to eliminate the global oscillation associated with the p-modes. For that we applied the FFT technique and removed oscillations with phase speeds above the local sound speed.

\begin{figure}[!h]
\center
\includegraphics[width=15.5cm]{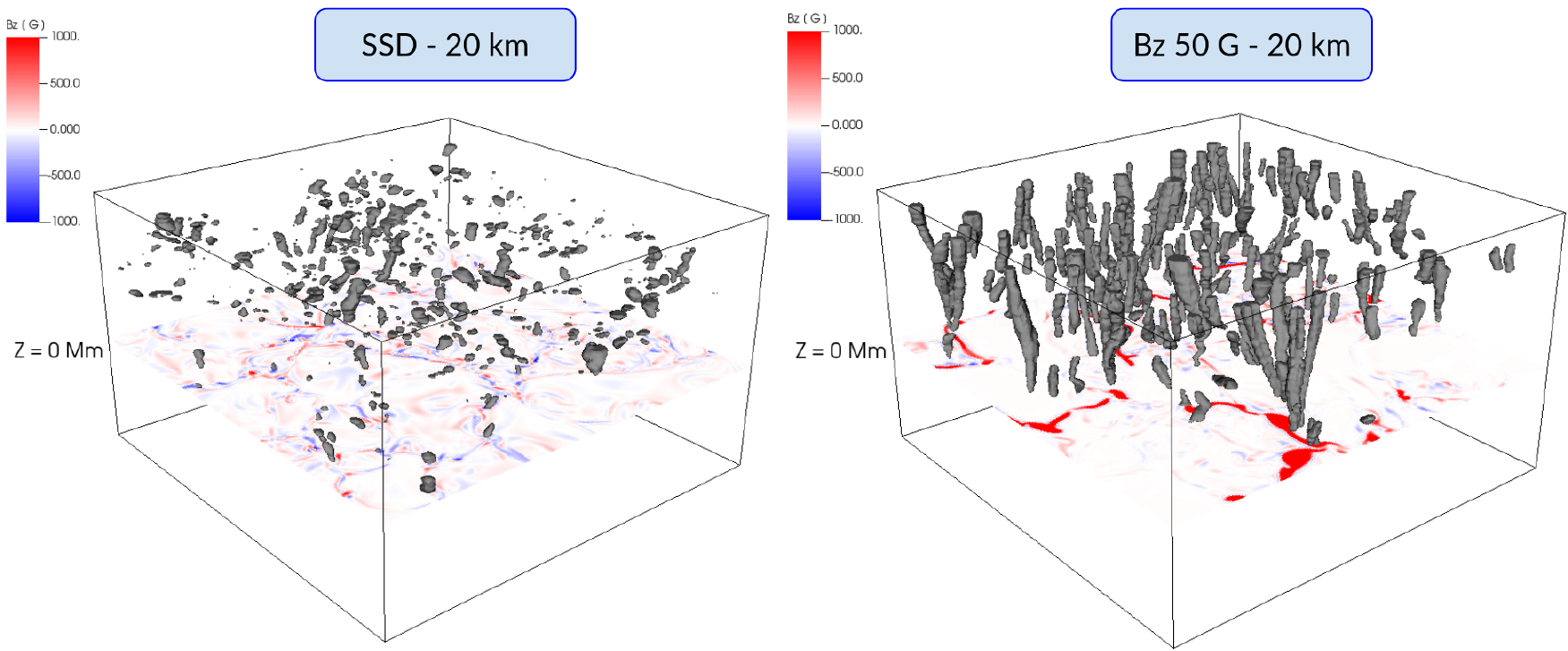}
\caption{\label{fig2} $3$D rendering of the detected vortices over the simulation box for the SSD (left) and $50$ G (right) simulations at $20$ km horizontal spatial resolution. The $2$D colormap indicates the vertical magnetic field at Z$=0$ Mm.}
\end{figure}

Fig.~\ref{fig2} shows two examples of vortex detections over the entire simulation domain. The gray contours represent the vortex detections that the code has made. For the SSD simulation (left panel), it can be seen that there are no large and coherent vortex structures and they are widely distributed throughout the domain. In contrast, in the $50$ G simulation (right panel), structures with high spatial coherence extending in the vertical direction from the photosphere to the top boundary of the simulation box are observed. This kind of behavior is due to the presence of stronger and more vertical magnetic fields structures in the $50$ G simulations and have been previously reported in other studies \cite{tziotziou23}. The area covered by vortices increases with height in all the models, reaching up to $1-1.5 \%$ in the highest resolution SSD models, and up to $6 - 7 \%$ in the unipolar models.  

\section{Results: Statistical properties}

\subsection{Number and size}

We determined the mean number of vortices as a function of height. In all of the models, we observed a minimum in the number of detections around $200$ km. This minimum may be related to the change in the energy propagation mechanism. Up to $200$ km, the velocity field is dominated by overshooting convective flows while as we move upward the velocity field starts to be dominated by waves \cite{nordlund09}. From approximately $200$ km height, the number of detections increases in larger or lesser extent depending on the considered model. In general, we have observed that both with increasing magnetic field and spatial resolution, the number of vortices increases significantly.

The radius of vortices are in the range between $20$ and $100$ km in all of our models. Similar values can be found in previous works in numerical simulations \cite{yadav21, canivete23}. The main effect associated with increasing the spatial resolution for the same magnetic field configuration is a decrease in the vortex sizes. At higher resolutions smaller vortices appear, which were not present at lower resolutions, producing the decrease in the average radius. The $5$ and $10$ km SSD simulations exhibit similar average vortex sizes. Thus, the higher resolution simulation may be reaching a saturation in the size of vortices. Including a magnetic field tends to shape the structure of vortices by increasing the average radius with height. This can be an effect of the magnetic flux tubes opening up with height.

\subsection{Temperature profiles}

To study the importance of vortices as energy propagation channels, we have analyzed the mean temperature profiles of these structures. The left panel of Fig.~\ref{fig3} clearly shows that the temperature in vortices (dashed black line) is always higher than the solar average (solid black line). This result is independent of the magnetic field configuration and spatial resolution. 

We observed that spatial resolution does not affect the temperature profiles of vortices. Increasing the magnetic field leads to the reduction of the difference between the vortex temperature profiles and the solar average.

We compared these temperature profiles with the temperature profiles of granules and intergranules. For this, we computed a mask for granules and intergranules at Z$=0$ Mm, applying it at all heights. In this way, we observed that the temperature profile of the intergranules fits very well with that of the vortices up to about $200$ km. This is expected since vortices are essentially located in intergranules regions. Temperatures in both structures increase with respect to the solar average up to about $200$ km because of reverse granulation. Higher up, the temperature of the vortices is above that of the intergranules.

\subsection{Heating terms}

In order to understand why vortices are always hotter than their surroundings, we computed the mean heating rates produced by viscosity, magnetic diffusivity and ambipolar diffusion, see \cite{khomenko24} for the details of the computation. The only term that is actually physical is ambipolar diffusion, while the other two terms are of numerical nature. 

\begin{figure}[!t]
\center
\includegraphics[width=6.7cm]{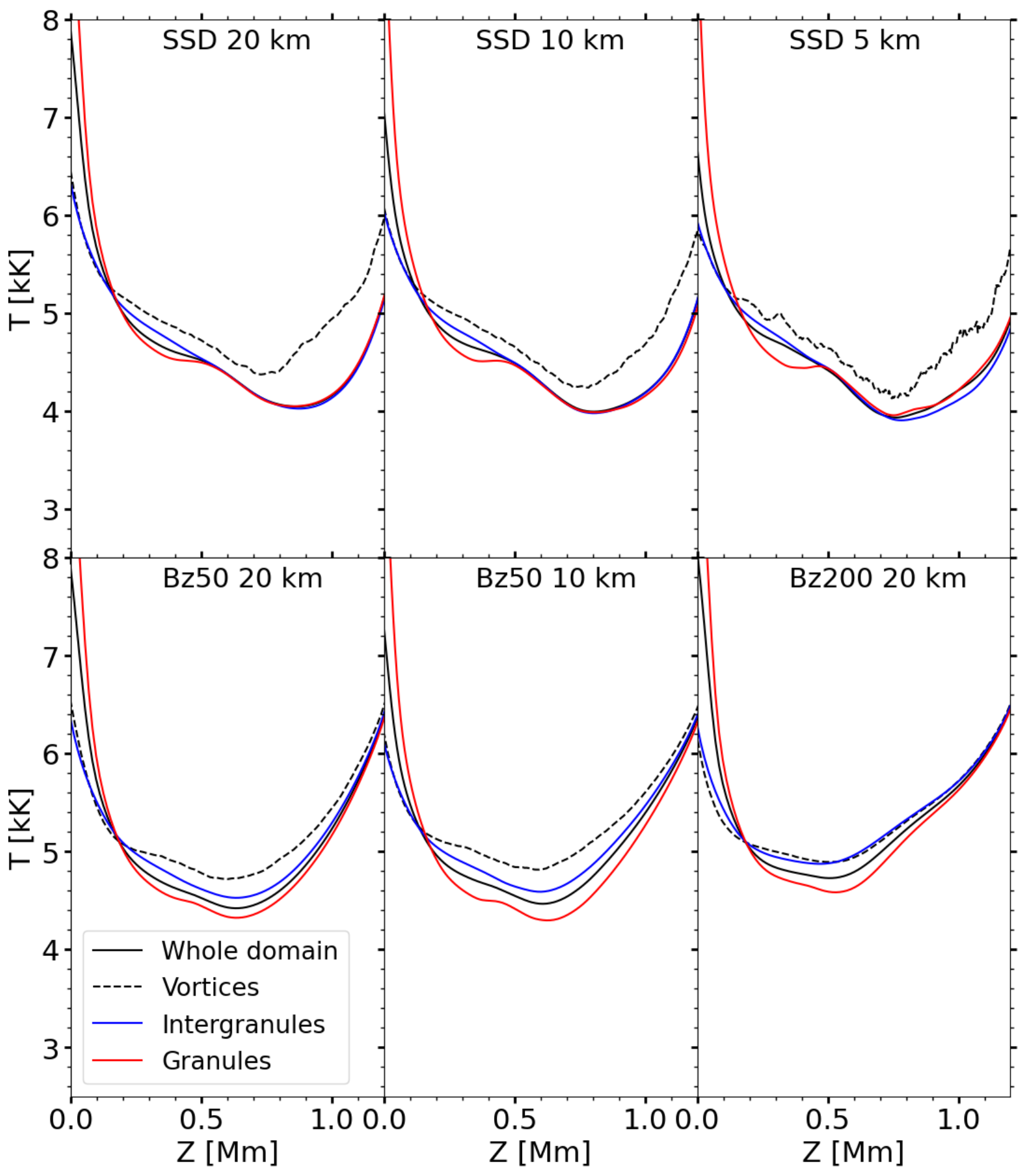} ~
\includegraphics[width=8.cm]{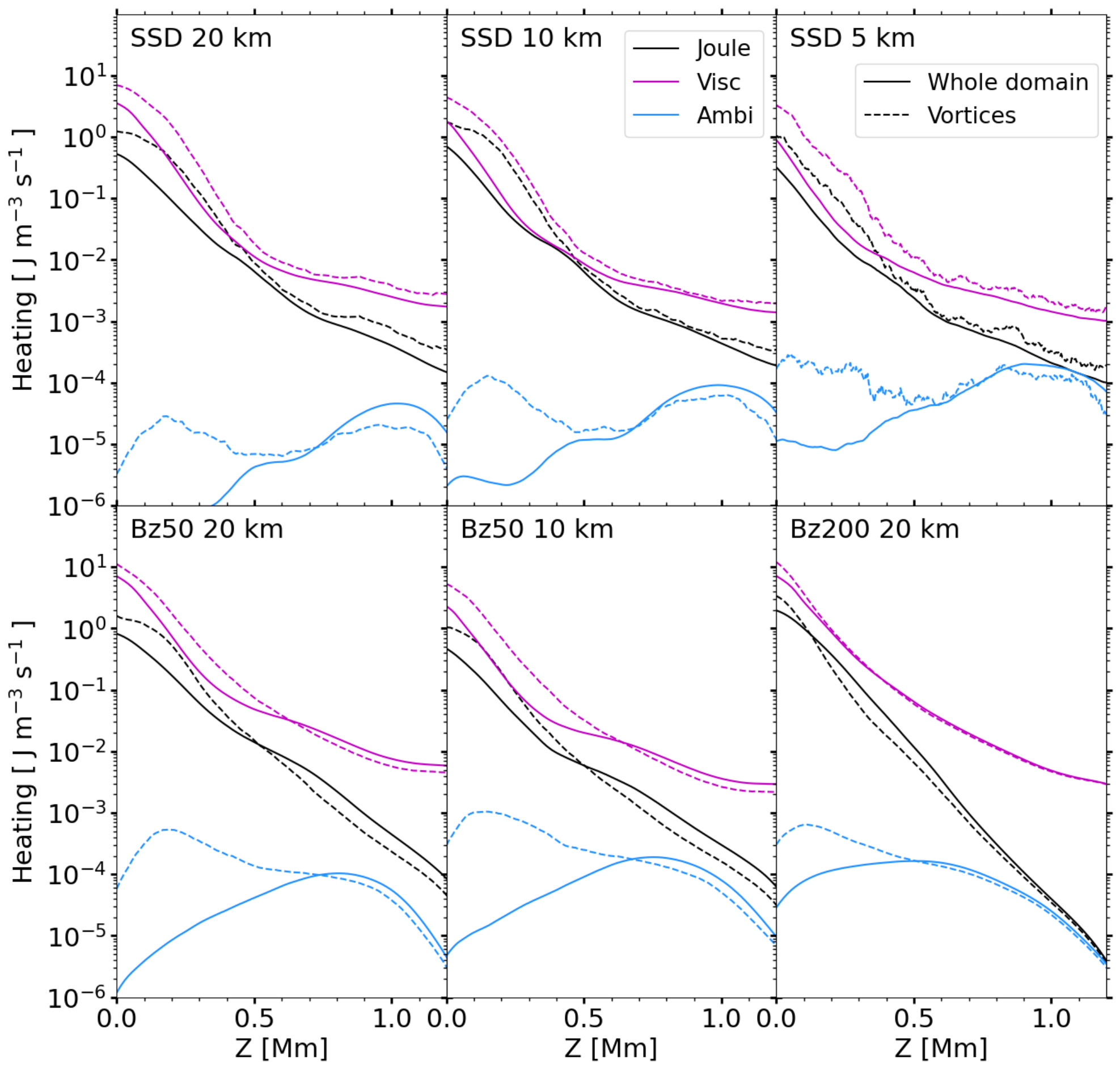}
\caption{\label{fig3} Left: mean temperatures in the whole domain (solid black), vortices (dashed black), intergranules (blue) and granules (red) as a function of height for all the models. Right: mean heating rates in vortices (dashed), in the whole domain (solid) as a function of height. Black: numerical magnetic diffusivity; magenta: viscosity; blue: ambipolar diffusion.}
\end{figure}

The results in the right panels of Fig.~\ref{fig3} demonstrate that viscous heating is the dominant term, followed by magnetic diffusivity and finally ambipolar diffusion. This result is independent of the magnetic field model and spatial resolution. Vortices have larger contributions to these terms than the solar average up to about $600$ km. Above this height, the contribution of vortices becomes similar to those of the average. In the $200$ G model, the contribution of vortices to viscosity and magnetic diffusivity never exceeds the mean values. 

Looking back at the temperature profiles (left panel of Fig.~\ref{fig3}), we observed that the largest temperature differences were located around $600$ km for the unipolar models and at higher layers for the SSD models. At these heights, the heating rates are similar to those of the solar average. Therefore, we speculate that vortices are able to dissipate energy in the lower layers of the atmosphere, where the heating rates are significantly higher than the average. Then, the hot plasma is transported by some mechanism to the upper layers producing this temperature increase in the profiles.

\section{Conclusions}

We carried out the detection and analysis of vortices in numerical 3D magnetoconvection simulations for different magnetic field configurations and spatial resolutions. We have applied the SWIRL code for vortex detection, validating its performance. We have found that different magnetic field configurations and spatial resolutions significantly affect the number and size of detected vortices.

We have clearly determined that the temperature of the vortices is always higher than that of the surroundings for all models. Thus, we confirmed the importance of these structures for the solar atmospheric heating. We found that vortices follow the same temperature profile as intergranules up to about $200$ km. Thereafter, vortices continue to be hotter than the surroundings. We analyzed different heating mechanisms to understand the increased temperature in vortices. We focused on viscosity, magnetic diffusivity and ambipolar diffusion, confirming that viscosity is the dominant dissipation mechanism at all heights. Vortices have higher contributions to these terms than the average only in the lower layers, up to about $600$ km. To account for the higher temperatures in vortices, we suggest that vortices dissipate energy in the photosphere and transports the heated plasma to the upper layers. 

\vspace{-0.1cm}
%
%
\small  
%
\section*{Acknowledgments}   
%

Financial support from grant PID2021-127487NB-I00, funded by MCIN/AEI/ 10.13039/501100011033 and by “ERDF A way of making Europe”, is gratefully acknowledged. TF acknowledges grant RYC2020-030307-I funded by MCIN/AEI/ 10.13039/501100011033 and by “ESF Investing in your future” and grant CNS2023-145233 funded by MICIU/AEI/10.13039/501100011033 and by “European Union NextGenerationEU/PRTR”. The authors thankfully acknowledge the technical expertise and assistance provided by the Spanish Supercomputing Network and the computer resources used: LaPalma Supercomputer, located at the Instituto de Astrofísica de Canarias, and MareNostrum based in Barcelona/Spain. The simulations analyzed in this work required about 15 millions of CPU/hours.

\vspace{-0.4cm}

%

%
\end{document}